# scientific reports

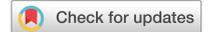

OPEN

# Ultra-broadband local active noise control with remote acoustic sensing

Tong Xiao✉, Xiaojun Qiu & Benjamin Halkon

One enduring challenge for controlling high frequency sound in local active noise control (ANC) systems is to obtain the acoustic signal at the specific location to be controlled. In some applications such as in ANC headrest systems, it is not practical to install error microphones in a person's ears to provide the user a quiet or optimally acoustically controlled environment. Many virtual error sensing approaches have been proposed to estimate the acoustic signal remotely with the current state-of-the-art method using an array of four microphones and a head tracking system to yield sound reduction up to 1 kHz for a single sound source. In the work reported in this paper, a novel approach of incorporating remote acoustic sensing using a laser Doppler vibrometer into an ANC headrest system is investigated. In this "virtual ANC headphone" system, a lightweight retro-reflective membrane pick-up is mounted in each synthetic ear of a head and torso simulator to determine the sound in the ear in real-time with minimal invasiveness. The membrane design and the effects of its location on the system performance are explored, the noise spectra in the ears without and with ANC for a variety of relevant primary sound fields are reported, and the performance of the system during head movements is demonstrated. The test results show that at least 10 dB sound attenuation can be realised in the ears over an extended frequency range (from 500 Hz to 6 kHz) under a complex sound field and for several common types of synthesised environmental noise, even in the presence of head motion.

Long-term exposure to either occupational or environmental noise can lead to a series of diseases, both auditory[1] and non-auditory[2,3]. Global active noise control (ANC) systems aim to reduce the undesired sound in a large environment, but require a large number of control loudspeakers, making them impractical in many applications[4,5]. By contrast, local ANC systems aim to reduce the sound only at specific (local) positions, often around a listener's ears, with the most common example being increasingly ubiquitous personal ANC headphones. In such solutions, the cushion and shell structure of the earcups provide passive sound attenuation in the mid to high frequency range (generally above 1 kHz) for auditory comfort[6]. In parallel, the active control component uses integrated speakers and microphones to produce anti-noise signals to attenuate sound in the lower frequency range (generally below about 1 kHz)[7]. ANC headphones are commonly used by passengers and aircrew in aircrafts, where the cabin noise during long haul flights is known to be detrimental to health and wellbeing[3,8]. However, longer-term use of such solutions can cause discomfort and/or fatigue because of the requirement to create a sealed volume around the ear which requires extra earcup clamping pressure. Solutions that can deliver sound reduction performance on par with that of earmuffs-based ANC headphones, but without the need to wear anything, would have value in many scenarios such as for machinery or equipment operators, for drivers of vehicles and for people working in open plan offices.

A substantial amount of effort has been made to move the required anti-noise loudspeakers (commonly denoted as secondary loudspeakers) and error sensors of ANC systems away from the user's ears and head while still realising effective noise reduction[9–14]. Indeed in an ANC headrest system (also known as an active headrest in some literature), both the secondary loudspeakers and the error microphones can be installed within the seat to reduce the undesirable sound (commonly denoted as primary sound) at both of user's ears[9]. Due to the difference between the sound pressure measured at these *remote* microphone locations and that at the user's ears, such a solution cannot guarantee sufficient sound reduction, particularly in the higher frequency range. Since the first proposal of ANC headrest technology in 1953[9], many virtual error sensing algorithms have been proposed which estimate the sound pressure at the ears based on the signals obtained from physical microphones positioned at alternative remote locations[15–20]. The upper-frequency limit for effective sound control remains relatively low

Centre for Audio, Acoustics and Vibration, University of Technology Sydney, Sydney, Australia. ✉email: Tong.Xiao@student.uts.edu.au

with a recently proposed system consisting of an array of four microphones and a head tracking system still only achieving sound reduction up to 1 kHz for a single sound source[14,20,21]. While this may be sufficient for certain low frequency ambient or machine induced noise, it is less effective for speech or other higher pitch sounds. In addition, the number of microphones required can be significant when using such a technique for real-world applications; this is not desirable in many applications.

Compared with ANC headphones, which are now well-developed and are increasingly popular and accessible products in the consumer electronics marketplace, the ANC headrest is an alternative solution which aims to provide a quiet environment for a user *without* the use of passive sound attenuating materials (i.e. the earcup or earbud). However, due to the physical limitation of moving the required components and absorbing materials away from the user, the development of ANC headrests has seen little progress despite ongoing research and development over several decades. It should be emphasised that such ANC headrests are not necessarily comparable to other sound control/attenuation devices, such as earplugs[22], ANC earphones with inserts[23] or modified hearing aids with a potential ANC functionality[24]. The fundamental principle of and motivation for an ANC headrest system is not to impose any disturbance on the user, whereas all of these alternative devices do.

There are other alternative ANC systems have been previously proposed. For example, the use of optical microphones based on optical fibres have been explored to replace traditional microphones in an ANC system deployed in an unfavourable environment, such as near a magnetic resonance imaging scanner[25]. Despite some effort on miniaturisation of the fibre-optic microphone, such a solution can still not be considered non-invasive, however, due to the physical presence of the optical fibres. In addition, while alternative MEMS microphones can be made extremely small, these devices require various accessories, such as batteries for power and wireless modules for data transmission, ultimately resulting in a solution with a considerable installation footprint. The challenge is clearly consistent with that for traditional condenser microphone-based solutions; the use of a direct *fully remote* acoustic sensing method in an ANC system would clearly offer advantages over these various other possibilities.

In this paper, a remote acoustic sensing approach using a laser Doppler vibrometer[26] (LDV) and a small membrane "pick-up" is examined in a real-time ANC headrest system. This arrangement includes a retro-reflective film as the membrane pick-up located at the cavum concha of a user's ear with an error-sensing LDV being positioned at a location remote from the user. LDVs typically have very high sensitivity with commercially available instruments able to resolve vibration displacements down to pm and velocities down to nm/s resolution. The membrane pick-up can be designed to be small and lightweight and have a wide dynamic range. Furthermore, being retro-reflective, the membrane can tolerate a wide range of inbound laser beam incidence angles, making this combination suitable even when the user's head moves (in combination with measurement point tracking as also described). Importantly, however, such a remote acoustic sensing approach can be highly attractive in an ANC application because the bulk of the signal processing components are located away from the user. The only minor encumberment on the user, to yield a practically realisable solution, is that they must wear a small optical pick-up, of no more mass nor volume than a typical piece of ear jewellery, on or ideally close to each ear canal.

Since the system does not include any largely intrusive actuators, sensors or bulky materials around the user's ears, it is described as a "virtual ANC headphone" in this paper. It neither installs any bulky error microphones with accessories on or near to the user's head, nor uses a large number of microphone arrays with certain virtual sensing algorithm methodologies to estimate the sound pressure at the ears based on measurements from elsewhere. As will be shown in the following section, this virtual ANC headphone system has significantly better performance than any other virtual error sensing solutions in the published literature thus far. While there remains further work to yield a commercially viable practical version of the solution, it is proposed that the technical benefits justify its proposal and continued investigation.

## System design and results

The system design and results of the subsequent experimental investigation are organised into five subsections. Initially, the system design of the virtual ANC headphone is described. Subsequently, the location of the membrane for the best control performance is examined. Thirdly, ANC performance in the presence of broadband grey noise is determined with the system implemented on a head and torso simulator (HATS). Penultimately, system performance is then evaluated for different kinds of synthesised real-world environmental noise signals. Ultimately, the use of a simple measurement location tracking system is incorporated to enable inevitable user head motion to be tolerated.

**Virtual ANC headphone system design.** A schematic showing the proposed system components and their arrangement is shown in Fig. 1a. Two secondary loudspeakers are placed behind a user's head (as they would be if integrated into a headrest), one at either side to control the primary sound from the surrounding environment at each ear and to thereby place the user in a quieter environment. An LDV is used to determine the acoustical signal at the ear canal entrance by measuring the surface vibration of a small, lightweight and retro-reflective membrane pick-up located nearby. While Fig. 1a shows two inbound laser beams, one to each ear, a single-ear solution is considered and described herein for the sake of brevity and clarity but with no loss of generality for the two-ear equivalent.

For ANC systems, a quiet zone is defined as a region in which more than 10 dB sound attenuation is achieved, with the zone size being about a tenth of the wavelength of the sound in a diffuse sound field[4]. When the membrane is placed close to the ear canal, such a quiet zone can be created around it, thereby reducing the sound propagating to the tympanic membrane (eardrum). The two secondary loudspeakers presented here were placed 0.44 m apart with an azimuth angle of 45 degrees pointing to the user as shown in Fig. 1b. The controller takes the surface vibration velocity of the membrane from an LDV as the error signal for the adaptive control, the details of which can be found in the *Methods—Noise control algorithm* subsection.

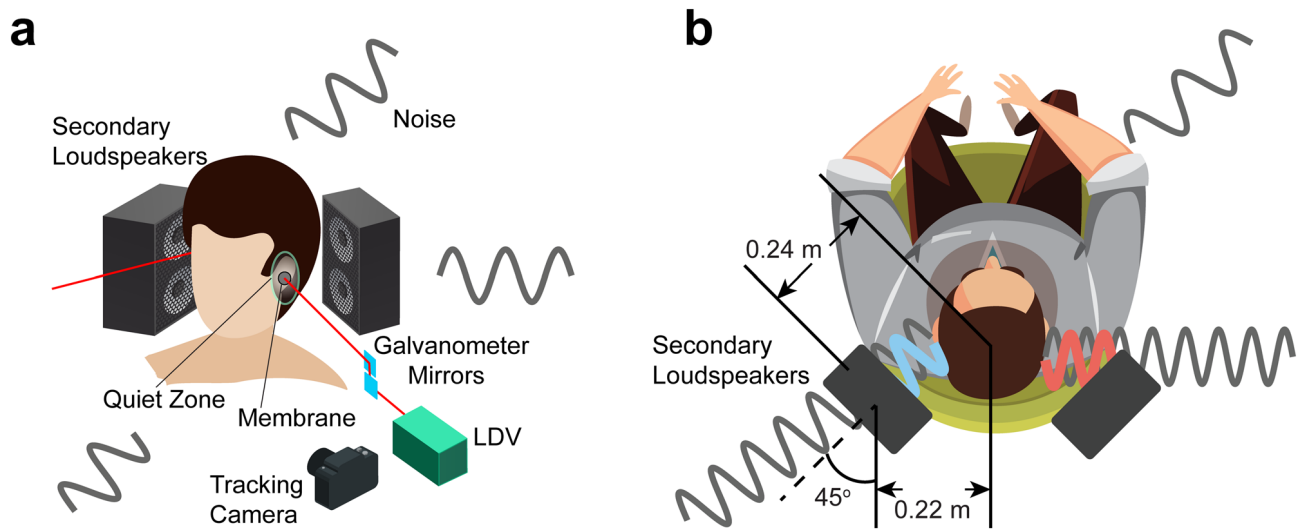

**Figure 1.** A virtual ANC headphone. (**a**) A quiet zone is formed in each ear by using a nearby secondary loudspeaker pair to reduce the sound in the ear, the required error-signal being determined from an LDV measurement of the vibration of a small membrane pick-up located close to the ear canal. Movement of the user is accommodated by a camera-based tracking system, which actively controls the galvanometer-driven mirrors to steer the laser beam and maintain its position on the membrane. (**b**) The locations of the secondary loudspeakers. Each secondary loudspeaker generates anti-noise signals through the ANC controller (not shown).

Normal head movements can be accommodated by a relatively straightforward camera-based tracking system, outlined in Fig. 1a, which actively controls a pair of orthogonal, galvanometer-driven mirrors to maintain the probe laser beam incidence on the centre of the membrane. Through the application of a bespoke image processing algorithm, the LDV can thereby remotely obtain the acoustical error signal in real-time.

The experimental setup is presented in Fig. 2a. The experiment was performed in a quiet room with a background sound pressure level of 38.5 dBA (A-weighted SPL, dB re. 20 µPa). A head and torso simulator (HATS; Brüel and Kjær Type 4128-C) with right and left ear simulators was used to measure the sound that would be experienced at the eardrums in a user's ears. Figure 2b shows the design and the configuration of the membrane pick-up used in this system. The pick-up consists of a piece of retro-reflective film (3 M—Scotchlite Sheeting 7610[27]), 0.1 mm in thickness, stretched over and adhered to a short, enclosed polymeric cylindrical tube with a diameter of 9.2 mm, a depth of 4.6 mm and a mass of approximately 0.2 g. The resulting combination is therefore as minimally invasive as practically possible in terms of size and mass. The film was used as the membrane so as to maximise the backscattered optical signal in relation to the inbound laser beam, irrespective of a non-normal beam incidence, this being advantageous in the presence of inevitable head movements. The membrane works similarly to a microphone diaphragm, converting the acoustical pressure induced mechanical vibration ultimately to an electrical signal. However, in this case, there are neither any electronic components inside (e.g., a preamplifier to process the measured signal), nor the need for wiring for signal transmission. Instead, signal conditioning and conversion are completed remotely in the LDV opto-electronics. Detailed parameters for the retro-reflective material and the frequency response of the membrane pick-up have been determined and can be found in Supplementary Fig. S1 and Supplementary Table S1.

The data acquisition system is at a remote location along with the LDV in the proposed arrangement. The LDV (Polytec PDV-100) has a measurable frequency range from 20 Hz to 22 kHz. The LDV was mounted on a tripod, vibration isolated from the HATS and the loudspeakers (Genelec 8010A). The sampling rate of the ANC controller (Antysound TigerANC WIFI-Q) was set to 32 kHz, and the filter lengths for both primary and secondary paths were set to 1024 taps. It should be noted that the adaptive control algorithm simply took the measured membrane velocity signal directly and attempted to minimise it. While the velocity signal could potentially be converted into sound pressure by some means, this was not necessary—the outcome would be the same whether it be the raw signal or some derivative of it.

**Optimal placement of the membrane pick-up.** Although obvious to place the membrane pick-up as close to the ear canal as possible, it is not immediately clear which specific location/s were more feasible/optimal and what the ANC performance might be for each. Four possible pick-up locations are illustrated in Fig. 3, where location #1 is on the anterior notch of the pinna, location #2 is on the tragus, location #3 is in the cavum concha, and location #4 is on the lobule. The experiments were performed in the left synthetic ear of the HATS. Only one loudspeaker, located 0.6 m away directly to the rear of the HATS, is used here as the primary source. The primary source signal was a broadband grey noise with a customised Fletcher-Munson curve filter[28] from 500 Hz to 6 kHz (see Supplementary Fig. S2). The filter was applied here to yield a measured SPL with a flat frequency response inside the HATS. The overall SPL at the left tympanic membrane was 77.7 dB (re. 20 µPa—omitted hereafter for brevity) with ANC off.

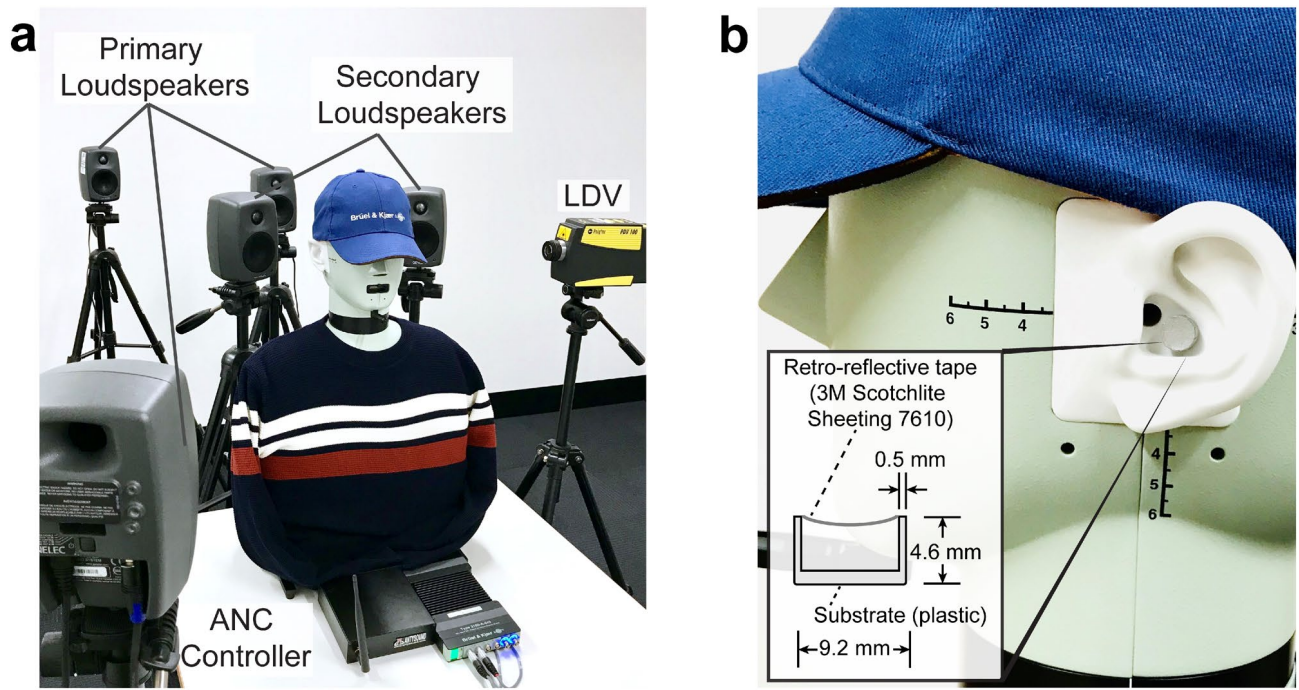

**Figure 2.** Experimental setup for a stationary HATS. (**a**) Two secondary loudspeakers were placed behind the HATS for sound control. Multiple primary loudspeakers (three shown) were located arbitrarily to simulate unwanted sound from different directions. The probe laser beam from the LDV was directed toward the membrane in the ear. (**b**) A membrane was placed close to the ear canal of the left synthetic ear of the HATS. The LDV remotely determines the surface velocity of the membrane as the error signal for the ANC controller.

With ANC on, the performances at locations #1 and #2 were similar with the resulting overall SPL being 69.2 dB and 70.9 dB, respectively. However, the sound reduction was only significant at frequencies below 4 kHz. The reason may be that the sound pressures measured at these two points are only similar to that at the ear canal below 4 kHz. Thus, the control performances at the two points are also limited up to 4 kHz. The sound reduction at location #3 was the best with an overall SPL of 63.5 dB when ANC was on. The overall SPL was reduced by 14.2 dB over the entire frequency range from 500 Hz to 6 kHz. Location #4, the lobule, was further away from the ear canal than any of the other selected locations. The effective frequency range of the sound reduction was only up to approximately 3 kHz with an approximately 6 dB *increase* in fact observed over the 5 to 6 kHz range. Based on the outcomes from this membrane location performance analysis, location #3 (the cavum concha) was identified as the optimal location for the membrane; in the remaining experimental investigations described herein, this is therefore the membrane position employed.

**Performance evaluation for broadband noise.** Figure 4 shows the measured noise spectra for each ear without and with ANC for three different primary sound field scenarios. Loudspeaker(s) driven with common signals were arranged to create increasingly complex surroundings with one or multiple reflectors. The signal used was again the broadband grey noise equivalent to that used to obtain the results presented in Fig. 3. All the test results were obtained by averaging over a 15-s data length. Figure 4a shows the setup where a single primary source was located 0.6 m away directly to the rear of the HATS to simulate the sound coming from a nearby source without considering any reflections from the surroundings. After enabling ANC, almost 15 dB attenuation was realised with the overall SPL being reduced from 78.1 dB to 63.8 dB and from 77.3 dB to 62.0 dB at the left and right ears respectively. This scenario is similar to that presented in the current state-of-the-art system[20], where the sound up to 1 kHz was controlled, albeit here the improvement achieved is over a much wider frequency range, up to 6 kHz. It is worth noting that the tests were still performed at each side separately instead of being taken simultaneously in this case.

Figure 4b shows the setup and results from a situation in which two primary loudspeakers were placed arbitrarily at two different locations. This can represent a situation when the user is close to a large rigid reflecting surface, such as a table or a wall. In this case, the acoustic signals from the original source and the reflector are coherent. Approximately 13 dB attenuation was obtained with the overall SPLs being reduced from 80.2 dB and 77.9 dB to 66.0 dB and 65.2 dB at the left and right ears, respectively. Figure 4c shows a more general situation where multiple reflectors exist. Four primary loudspeakers were arbitrarily positioned at various locations around the head to achieve this. Approximately 11 dB attenuation was obtained with the overall SPL reduced from 80.4 dB to 68.9 dB and from 80.1 dB to 69.4 dB at the left and right ear respectively. In all three of these example scenarios, the demonstrated system yielded a minimum 10 dB reduction across the entire 500 Hz to 6 kHz frequency range. It is worth noting that the placements of these primary sources were created arbitrarily, however, the control performances observed are expected to be similar for any other similar configuration.

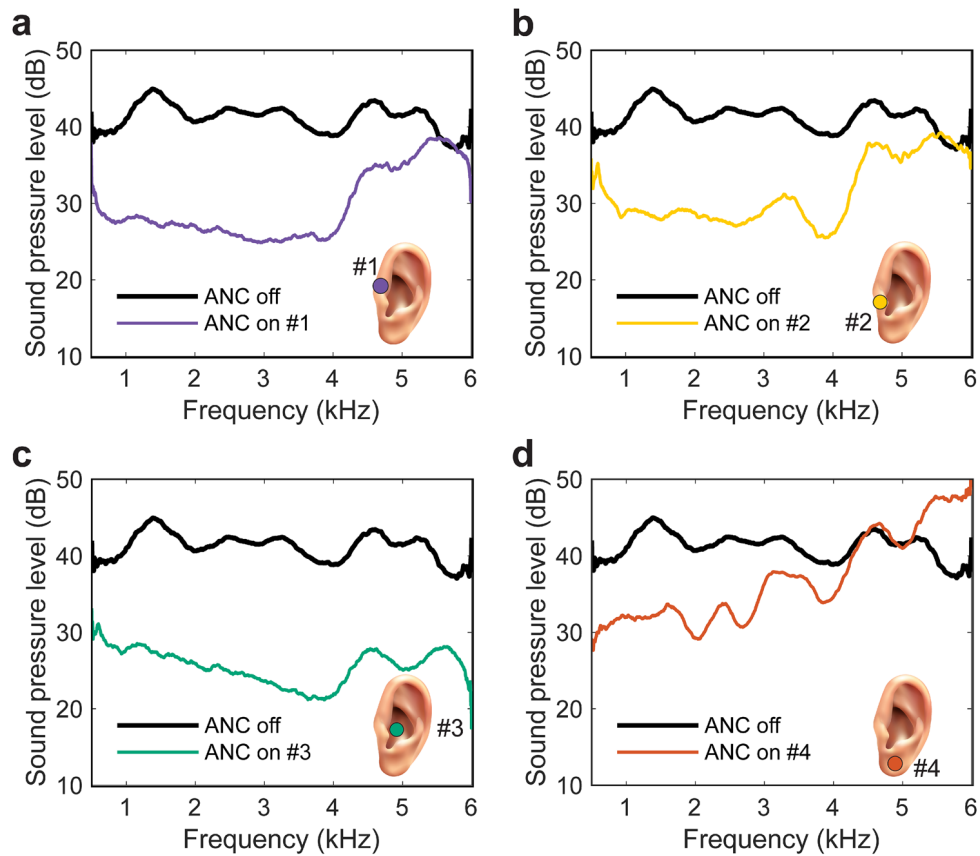

**Figure 3.** The SPLs (dB re. 20 µPa) measured from the left ear simulator of a HATS, simulating the sound that a user experiences at the left tympanic membrane with and without ANC, when the membrane was at (**a**) location #1—anterior notch; (**b**) location #2—tragus; (**c**) location #3—cavum concha; and (**d**) location #4—lobule of the HATS left synthetic ear.

**Performance evaluation for synthetic environmental noise.** To further demonstrate the capability of the proposed solution, performance in the presence of three different kinds of pre-recorded common environmental noise scenarios was assessed. Similar to the configuration implemented recently[20], the primary source was located about 1.2 m directly behind the HATS, with only one channel (right ear) being controlled. The three experiments were performed in a hemi-anechoic chamber. Firstly, a recording of aircraft interior noise[29] was used as the primary source signal. The 15-s signals observed by the HATS before and after ANC are shown in Fig. 5a with the corresponding spectra averaged over this duration also shown. The overall SPL was significantly reduced from 74.7 dB to 59.6 dB—a greater than 15 dB improvement. Secondly, an example of an aircraft flyby noise[30] was examined. Figure 5b shows the time-domain signal observed by the HATS of such non-stationary noise before and after ANC and the spectrum (averaged from 3 to 8 s only). Again, there was a significant reduction over the 500 Hz to 6 kHz range. Indeed, where the noise was the most pronounced, i.e. from 3 to 8 s, the overall SPL was reduced from about 82.1 dB to 61.6 dB—a greater than 20 dB sound attenuation. Lastly, a recording of a crowd of people talking was used as the primary source signal[31]. Figure 5c shows the 15-s time-domain and the frequency-domain signals before and after ANC again. The overall SPL was controlled from 75.5 to 59.8 dB; over 15 dB reduction was achieved. Table 1 summarises the averaged overall SPLs without and with control using the proposed system for these new scenarios, where 15–20 dB noise reduction up to 6 kHz can be achieved using the proposed system. The audio recordings before and after ANC can be experienced through Supplementary Movie 1. It is important to note that the current state-of-the-art virtual sensing ANC solution, with a quoted upper frequency performance of around 1 kHz, would not yield as impressive a performance as the virtual ANC headphone presented herein since, as can be observed in Fig. 5, the more significant frequency content in all three example signals primarily exists in the 2 to 4 kHz range.

**Performance evaluation in the presence of head motion.** A person is prone to exhibit continuous head motion, therefore, the probe laser beam from the LDV should be able to track the corresponding arbitrary motion of the membrane in the ears. Such tracking LDV solutions have been widely researched, developed and applied for numerous complex measurement tasks[26]; the scenario herein represents a further interesting application. A simple tracking system was therefore implemented to demonstrate the proof of concept. This bespoke camera-based tracking system is shown in Fig. 6 with specifications presented in the *Methods—Head tracking system* subsection. The scenario used here is the same as the one described in Fig. 4a, i.e. that with a single sound source immediately to the rear.

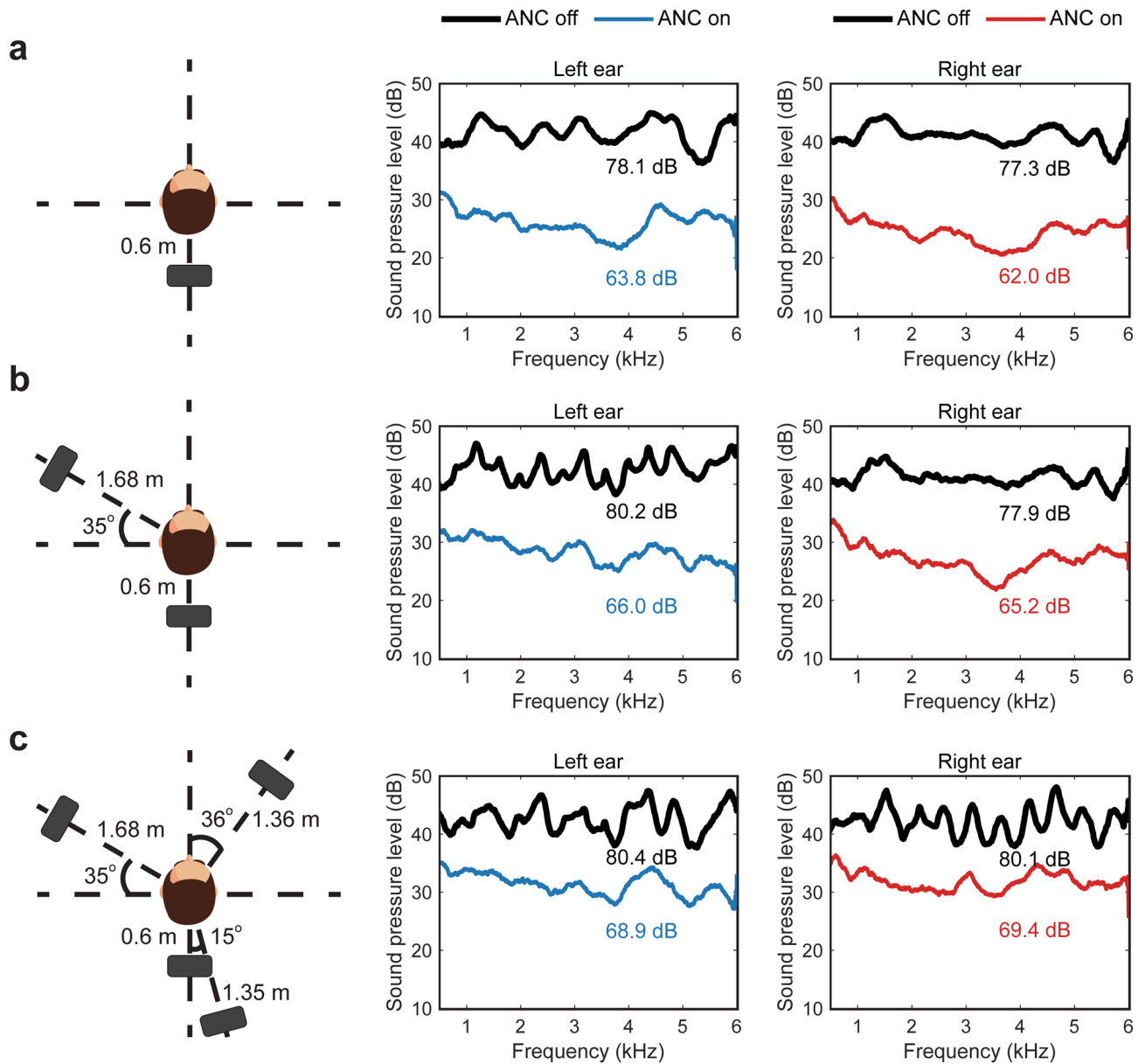

**Figure 4.** Three configurations of the primary loudspeakers and the corresponding SPL (dB re. 20 μPa) with and without ANC at both ears. (**a**) A single primary loudspeaker was used to simulate the sound from a single source nearby. (**b**) Two primary loudspeakers were used to simulate two sound sources nearby or a single sound source with a nearby reflecting surface. (**c**) Four primary loudspeakers were used to simulate sound from multiple directions, approximating a general case in practice.

The movement of a marker on the ear lobule of the HATS, as illustrated in Fig. 6c was determined by the image processing-based tracking system to maintain near-optimal laser beam incidence on the membrane and yield a useful error signal. Supplementary Figure S3 and the associated remarks present the effects of off-centre measurements and different laser beam incident angles on the system performance. Overall, the performance was not particularly sensitive to the precise location of the laser beam on the membrane, with it therefore deemed not necessary for the laser beam incidence to be precisely at the geometrical centre. With the laser beam slightly off-centre, ANC performance is maintained. Furthermore, the incidence angle of the laser beam did not affect performance significantly. With incidence at a quite remarkable 60 degrees, the LDV signal drops by around 5 dB, which, again, has minimal detrimental effect on the ANC performance. These characteristics have laid the foundation for the successful application of the tracking system to manage inevitable user head movements.

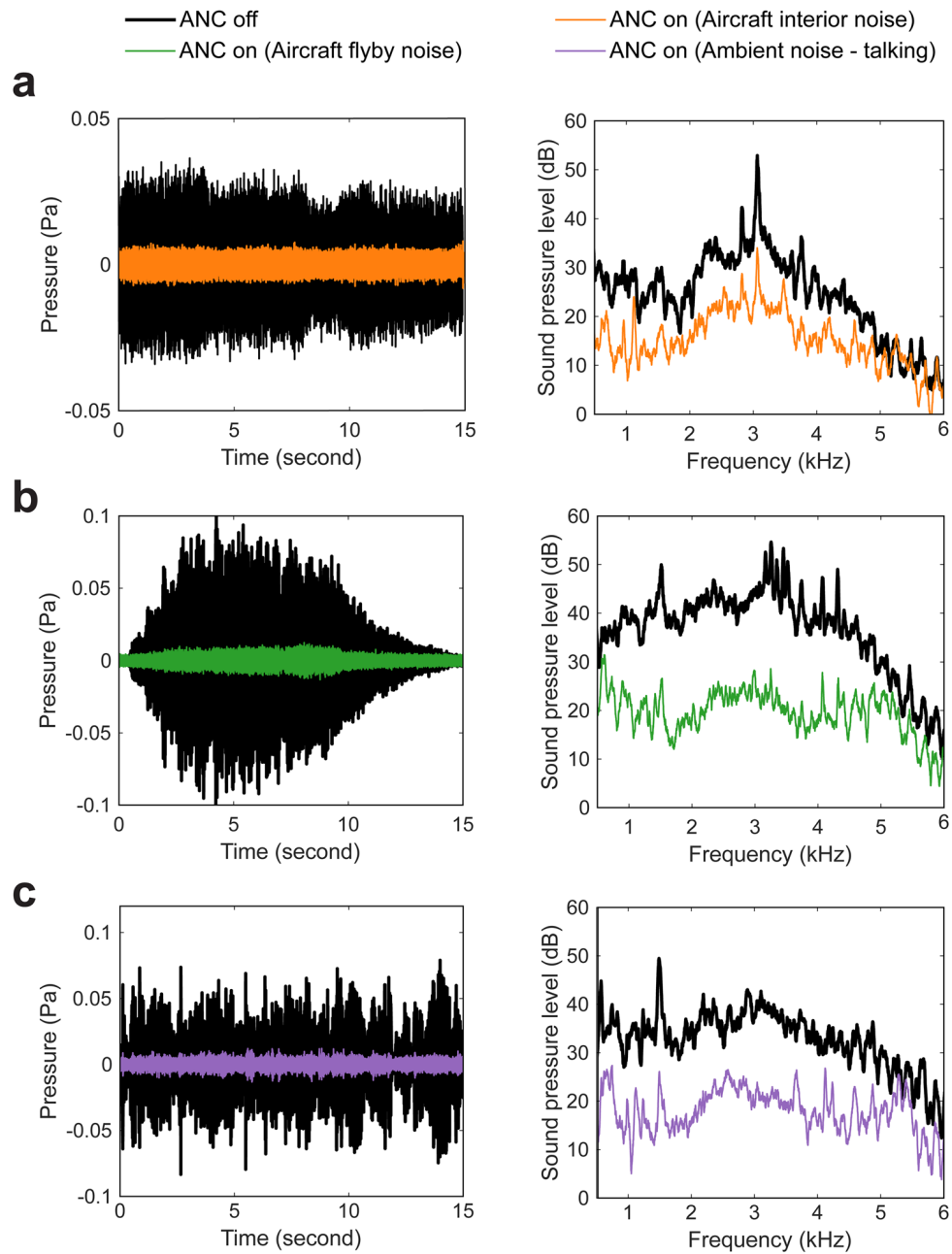

**Figure 5.** The time-domain signal observed by the HATS and the corresponding sound pressure level (dB re. 20 µPa) without and with ANC for (**a**) aircraft interior noise, (**b**) aircraft flyby noise and (**c**) ambient noise of people talking.

Figure 7 shows four control performances—when ANC is off (1) and on (2) for a *stationary* HATS and when ANC is on with the head tracking system disabled (3) and enabled (4) for a *moving* HATS. The movement of the HATS was implemented manually with a forward–backward movement used to simulate a person moving back and forth while seated. The maximum distance the HATS travelled in the Supplementary Movie 2 was approximately 0.08 m peak-to-peak with a maximum speed of about 0.04 m/s. Figure 7a shows the 15-s sample of the time-domain measurement for each case with the same configuration as in Fig. 4a. Figure 7b shows the corresponding averaged frequency spectrum for each case for the entire duration. Similar to the results previously presented in Fig. 4a, the total SPL was reduced from 81.1 to 64.1 dB over the frequency range from 500 Hz to 6 kHz range for the stationary situation.

When the HATS moved with ANC on but with tracking disabled, the head (therefore the membrane) moved away from the probe laser beam; the LDV signal thereby "dropped out" or made a vibration measurement not representative of the sound pressure at the ear. This can easily make the control system diverge and, as shown in Fig. 7b, the overall SPL in fact *increased* significantly from 81.1 to 99.5 dB. When the tracking system was enabled, the mirrors maintained the laser beam incidence on the membrane as the HATS moved. Thus, the LDV measurement remained valid for the adaptive control. As shown in Fig. 7b, the system reduced the sound from 81.1 to 70.4 dB over the entire frequency range. The control performance maintained at least a 10 dB reduction during the movement of the HATS, demonstrating the necessity of using a tracking system for the ANC system. Again, these audio recordings can be experienced in Supplementary Movie 2.

| Noise type (average duration) | Averaged overall SPL (dB) | | |
|---|---|---|---|
| | Without ANC | With ANC | Noise reduction |
| Aircraft interior (0–15 s) | 74.7 | 59.6 | 15.1 |
| Aircraft flyby (3–8 s) | 82.1 | 61.6 | 20.5 |
| Ambient speech (0–15 s) | 75.5 | 59.8 | 15.7 |

**Table 1.** The averaged overall SPL without and with proposed ANC system for three types of synthetic example environmental primary noise.

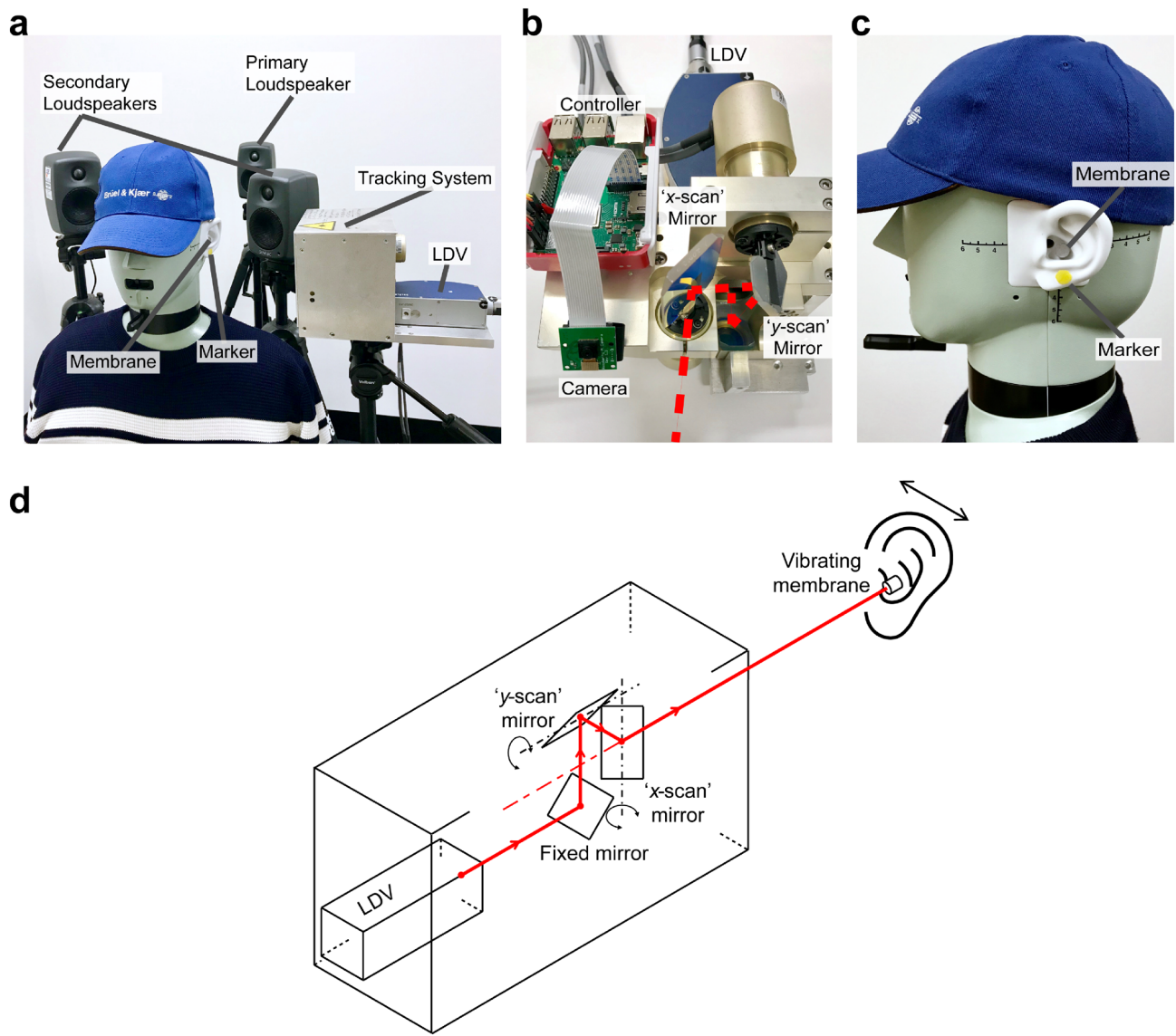

**Figure 6.** (**a**) Configuration of the head tracking system with a single primary loudspeaker. The tracking system and the LDV are placed to the left side of the head. (**b**) The construction of the tracking system with a pan and a tilt mirror for steering the laser beam. The camera is attached to the controller for the target object tracking. (**c**) A yellow marker is placed below the membrane on the ear lobule as the target object. (**d**) Schematic of the camera-based tracking system showing the laser beam path from the scanning LDV.

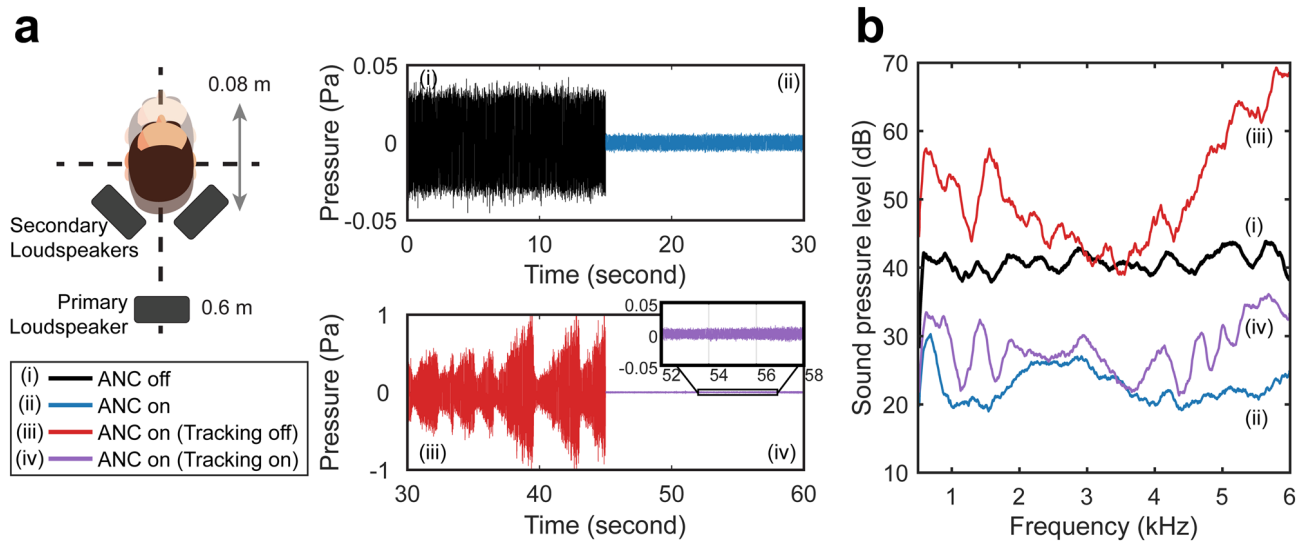

**Figure 7.** ANC performance with the developed head tracking system. (**a**) Four 15-s samples of the time-domain signal observed by the HATS. The upper 30 s duration shows the sound pressure with ANC off and on for the stationary situation, while the lower 30 s duration shows the sound pressure with ANC on with the tracking system off and on for a moving HATS. (**b**) The corresponding sound pressure level (dB re. 20 μPa) of the four signals.

## Discussions

Like many other systems, the demonstrated solution also faces certain limitations. Particularly, while the demonstrated system using the remote acoustic sensing approach can achieve an ultra-broadband control, the cost of the required LDVs can be high. However, it is possible that these can ultimately be made smaller and at a lower cost[32] with the entire proposed system thereby being designed to be sufficiently compact and low cost to be used in headrests for example in airplanes or in (driverless) automotive applications in the future.

Some further limitations of the solution are also acknowledged. Firstly, the performance of such a virtual ANC headphone (or ANC headrests in general) is still inferior when compared to that of ANC headphones. In particular, sound attenuation achieved in the higher frequency range by active control is below that which can be simply achieved through passive control, i.e. earmuffs, with these often delivering over 30 dB reduction[33,34]. However, such comparison is unfair since the aim of an ANC headrest system is to eliminate the use of the passive attenuation materials which deliver such reduction.

Secondly, while multiple primary sources were used to simulate unwanted sound from multiple, arbitrary directions, the reference signals for the ANC controller were taken directly from these loudspeaker signals; in a real-world situation, this would clearly not be possible. The reason for taking the reference signal directly from the primary source was to focus on the proposed remote acoustic sensing approach. Future developments include incorporating one or multiple actual reference sensors into the system. The locations of these reference signal sensors are less constrained than those of the error signal sensors. However, they should still be close to the entire system, including to the secondary sources. The constraint of the possible locations would affect the control performance and this remains a topic to be further studied in the ANC community.

Thirdly, the head tracking system shown for illustration purposes was only capable of tracking two-dimensional motions. A more robust head tracking system with a higher frame rate camera and auto-focus could be implemented to accommodate faster, three-dimensional head movements in the future. Lastly, it should be noted that the laser type used in the experiment was the Class 2 Helium–Neon 633 nm. While eye-safe, it may cause an undesirable effect and distract/dazzle the user if visualised, even briefly. Alternatively, invisible infrared based LDVs could be used in future developments.

## Methods

**Noise control algorithm.** The ANC controller uses the feedforward structure with the filtered-x least mean square (FxLMS) algorithm[35]. The block diagram of the algorithm is shown in Fig. 8. Subscripts $L$ and $R$ are used in place of "Left" and "Right". For each side of the ear, the reference signal $x(n)$ was taken from the primary source for the adaptive ANC controller to calculate the control signal $y(n)$ for the secondary loudspeaker. An LDV measures the error signal $e(n)$, which can be represented as

$$e(n) = p(n) + \mathbf{w}^T \hat{\mathbf{r}}(n) \quad (1)$$

where $\mathbf{w}$ is the vector of controller coefficients and $\hat{\mathbf{r}}(n)$ is the estimated filtered reference signal vector, with $\hat{\mathbf{r}}(n) = \hat{s}(n) * \mathbf{x}(n)$, where $\hat{s}(n)$ is the impulse response of the estimated secondary path filter and $\mathbf{x}(n)$ is the reference signal vector. The controller coefficients are updated with

$$\mathbf{w}(n+1) = \mathbf{w}(n) - \mu \hat{\mathbf{r}}(n) e(n) \quad (2)$$

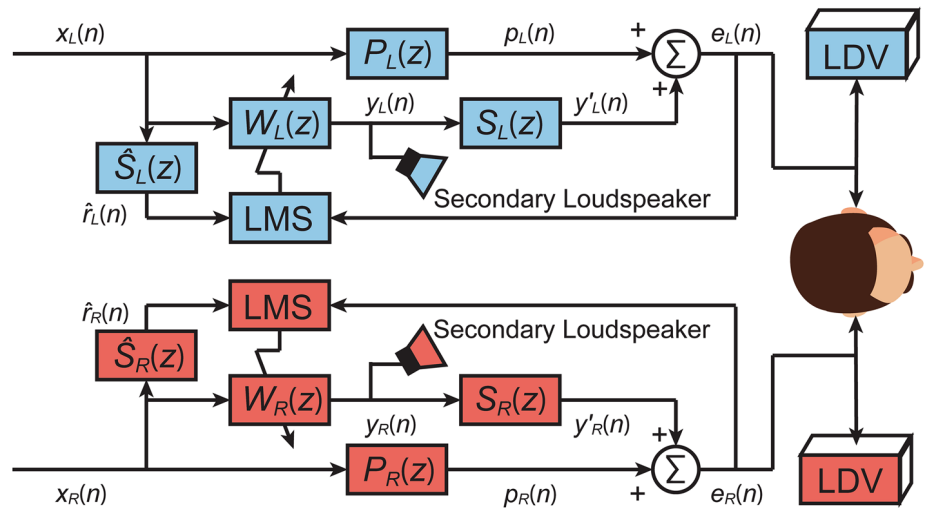

**Figure 8.** The block diagram of the adaptive active control algorithm for controlling the vibration velocity of the membrane measured by an LDV at each ear. At each side, the error signal is controlled separately by the corresponding secondary loudspeaker instead of being controlled simultaneously.

where $\mu$ is the convergence coefficient. The error signal $e(n)$ in this case is the membrane surface vibration velocity measurement from the LDV. The frequency plots of the error signals with ANC off and on are presented in the Supplementary Fig. S4 and associated remarks on the ANC performance at different SPLs.

**Head tracking system.** The LDV used with the tracking system was a Polytec NLV-2500-5 laser vibrometer, which was placed about 0.3 m away from the left ear of the HATS. The measurement sensitivity was set to 5 mm/s/V leading to a typical velocity resolution of 20 nm/s/√Hz[36]. The controller used in the demonstration for the object tracking was a Raspberry Pi 3B+, accompanied with a 30 fps, 5MP Omnivision 5647 camera module. A circular piece of yellow tape was adhered on to the ear lobule as a marker for object tracking purposes (shown in Fig. 6). After setting the laser beam to the centre of the membrane in the initial, stationary head stage, the controller detected any subsequent movement of the marker and therefore of the membrane. The movement of the membrane can then be translated into the movement of the marker in the controller. Revised galvanometer outputs, which are directly related to the position of the point-of-interest in two orthogonal directions ($x$ and $y$) in the plane of the motion, were derived. An MCP4725 digital-to-analogue converter was used to convert the digital signals from the Raspberry Pi to analogue voltage signals for the galvanometer mirror controller. The galvanometer mirrors were from GSI Lumonics with a matching, tuned for position, precision driver, MiniSAX. As a result, the steering mirrors adjusted the laser beam path such that it remained on the membrane enabling the LDV to measure the membrane surface velocity as the error signal for the ANC system.

The recognition of the object was implemented through colour extraction in OpenCV. The marker was extracted using an adequate threshold of a series of colour images. The image becomes binary as

$$B(x,y) = \begin{cases} 1, & (I(x,y) \geq \lambda) \\ 0, & (I(x,y) < \lambda) \end{cases} \quad (3)$$

where $I(x,y)$ and $B(x,y)$ are the pixel value of the original image and the binarised image, respectively. $\lambda$ is the threshold for the chosen marker.

Due to the circular shape of the marker, the mass centre of the target object can be readily determined, this being the centre of the marker. Between adjacent frames, the pixel shift of this centre was used to obtain the moving velocity of the object $\mathbf{v}_p(x,y)$. This in turn can translate to the velocity of the target object in reality $\mathbf{v}_r(x,y)$, which is expressed as

$$\mathbf{v}_r(x,y)\big|_D = \beta \mathbf{v}_p(x,y) \quad (4)$$

where $\beta$ is a scaling factor for a given distance $D$ between the camera and the target object. The value of $\beta$ was determined during the setup and calibration stage, where the parameters of the tracking system, such as the distance $D$ and the offset between the marker and the membrane, were specified.

## Conclusions

The performance of an ANC headrest using a remote acousto-optic sensing approach, proposed to provide a significantly quieter environment for a user, is investigated. The remote sensing approach uses an LDV and a small, lightweight and retro-reflective membrane pick-up placed in the cavum concha of a user's ear, thereby producing as little disturbance as possible. The membrane design using a retro-reflective film was presented and analysed, and the effects of its location on the system performance were explored. The noise spectra in the ears without and with ANC for different primary sound fields and diverse kinds of environmental noise were reported. A simple head tracking system was also developed to maintain the control performance during any possible head movements from the user. The results show that more than 10 dB sound attenuation can be obtained for an ultra-broadband frequency range up to 6 kHz in the ears for multiple sound sources and various types of common environmental noise. Future work will include enhanced membrane material design, a more robust head tracking system and the incorporation of reference signal sensors in place of signals taken directly from the primary sources.

### Data availability
The data that supports the findings of this study are available from the authors on reasonable request, see author contributions for specific data sets.

### Acknowledgements
This research is supported by an Australian Government Research Training Program Scholarship.

### Author contributions
T.X. contributed to idea conception, experimental design and measurements, data analysis and writing of the manuscript. X.Q. initiated and supervised the study, secured project/scholarship funding, contributed to idea conception and data analysis. B.H. advised on fundamentals of LDV and identified the need for the finite, scattering element—the membrane. All authors prepared, discussed and reviewed the manuscript.

### Competing interests
The authors declare no competing interests.

### Additional information
**Supplementary information** is available for this paper at https://doi.org/10.1038/s41598-020-77614-w.

**Correspondence** and requests for materials should be addressed to T.X.

**Reprints and permissions information** is available at www.nature.com/reprints.

**Publisher's note** Springer Nature remains neutral with regard to jurisdictional claims in published maps and institutional affiliations.